# MR-BART: Multi-Rate Available Bandwidth Estimation in Real-Time

Mahboobeh Sedighizad, Babak Seyfe, *Member, IEEE*, Keivan Navaie, *Member, IEEE*

*Abstract*—In this paper, we propose Multi-Rate Bandwidth Available in Real Time (MR-BART) to estimate the end-to-end Available Bandwidth (AB) of a network path. The proposed scheme is an extension of the Bandwidth Available in Real Time (BART) which employs multi-rate (MR) probe packet sequences with Kalman filtering. Comparing to BART, we show that the proposed method is more robust and converges faster than that of BART and achieves a more AB accurate estimation. Furthermore, we analyze the estimation error in MR-BART and obtain analytical formula and empirical expression for the AB estimation error based on the system parameters.

## I. INTRODUCTION

ACCURATE estimation of the Available Bandwidth (AB) over a network path with a temporally variable cross-traffic is a challenging problem [1-3]. The available bandwidth is particularly important when the network provides service to delay sensitive subscribers. In such cases an accurate estimation of the available bandwidth has a crucial role in providing Quality-of-Service (QoS) guarantees to the users [4-6]. In practice, several users share the network bandwidth and under some circumstances the bandwidth demand is higher than that of the link capacity which causes network congestion. Network congestion results in degradation of some QoS parameters such as transmission delay and packet loss. Therefore, an accurate estimation of the available bandwidth at each time instant has an important role in designing more efficient network resource management schemes [7].

In practice, a perfect estimation of the available bandwidth between two arbitrary nodes in the network requires access to the temporal traffic information of each node along that network path, which is not always possible for end-users.

M. Sedighizad and B. Seyfe are with the Department of Electrical Engineering, Shahed University, Tehran, Iran, (e-mail: seyfe@shahed.ac.ir)
K. Navaie is with the Department of Electrical and Computer Engineering, Tarbiat Modares University, Tehran, Iran, he is also with the Broadband Communications and Wireless Systems (BCWS) Centre, Department of Systems and Computer Engineering, Carleton University, Ottawa, ON, Canada, K1S~5B6 (e-mail: keivan.navaie@ieee.org).

To tackle this issue, probing schemes are utilized. The basic idea of probing is to inject a sequence of packets namely *probe packets* with pre-defined inter-packet time interval into the network path. Hereafter, we refer to the sequence of probe packets as the *probing sequence*. In the receiver side, the inter-packet times of the probe packets are affected by the actual available bandwidth. The available bandwidth is then estimated utilizing the relationship between the inter-packet time dispersions at the two ends of the network path. If the probe packet transmission rate exceeds the available bit rate of the network path, the probe packets are backlogged at some intermediate nodes, resulting in an increased transmission delay; otherwise, the probe packets is received with no delay. The available bit rate is then estimated as the probing rate at the beginning of congestion [8]. The main components of the above mentioned methods consist of *the pattern of probing sequences* and the *available bandwidth estimation method*.

Various probing schemes are proposed in the literature to estimate the available bandwidth through the information available at the network edges. Trains of Packet Pairs (TOPP) [8], path load [9], path chirp [10], PathMon [11], Pathvar [12] Bandwidth Available in Real-Time (BART) [13], and [14] are instances of probing schemes. Most of the existing available bandwidth measurement techniques impose a large amount of extra traffic load because of injecting the probing sequences. Furthermore, they usually require a long observation interval to estimate the AB with an acceptable level of accuracy.

In this paper, we propose a method which is an extension of the BART and is based on injecting multi-rate (MR) probing sequence which we call MR-BART. Our proposed method then utilizes Kalman filtering (KF) for AB estimation. In MR-BART, in addition to the changing the probing rate from one probing sequence to another, we also alter the probing rate within each probing sequence. This technique enables us to obtain a rich set of observations by injecting each probing sequence. The observed set of data is then utilized by a Kalman filter to adaptively estimate the available bit rate.

We also analyze the behavior of the estimation error of MR-BART based on the system parameters. Based on this analysis, the impact of the probing sequence parameters on the accuracy of estimation is investigated.

We show that MR-BART obtains a more accurate AB estimation and lees sensitive to the initial state of the Kalman filter than that of BART. The increment of the accuracy of estimation comparing to BART is mainly due to the increasing the dimension input data of the Kalman filter caused by injecting multi-rate probe packets. Selecting multi-rate for any probing sequence causes a smoother AB estimation and prevents from sharp increments which sometimes occur in BART.

The proposed method also introduces a set of adjustable parameters which increases its applicability into different scenarios. Performance evaluation is conducted by utilizing a simulated network environment. We compare our proposed method with BART approach which has been shown in [13] that outperforms other methods in the literature. Simulation results show that using our proposed approach a higher level of accuracy is achieved rather than BART, whereby the extra traffic load because of probing sequence is equal to the BART.

The rest of the paper is organized as follows; in Section II we elaborate on the probing procedure and present the system model. Then in Section III, our proposed method for available bandwidth estimation is presented. The performance of the proposed method is studied in Section IV. We present the simulation results in Section V. The paper is then concluded in Section VI.

## II. AVAILABLE BANDWIDTH ESTIMATION IN A NETWORK PATH

A network path from the transmitter to the receiver consists of a number of nodes. A node itself consists of a queue connected to an input link and an output link. The queues in the nodes have infinite buffer length and first in first out (FIFO) is the serving discipline. A network path consists of $L$ links, each with capacity of $C_l$ (bits/s), $(l=1,2,\ldots,L)$. The link capacity, $C_l$, is determined based on the physical layer interfaces of the transmitter and the receiver. The temporal variations of $C_l$ is slow so that it can be assumed constant in time-scale of interest. For link $l$, a cross-traffic with time varying rate of $y_l$ (bits/s) is considered, therefore, the available (residual) bandwidth of link $l$, $A_l$, is

$$A_l = C_l - y_l. \quad (1)$$

For a network path consisting of $L$ links, $A$, is defined as

$$A = \min_l A_l. \quad (2)$$

As it is seen in (2), for a network path, the Available Bandwidth (AB), is mainly determined by the link which has the minimum residual bandwidth. The link with the minimum residual bandwidth is called the *bottleneck link*. Consider a fluid flow with a constant-rate cross-traffic $y$ that is transported through a single hop network with the link capacity $C$. In the probing scenario the transmitter injects probe packets with instantaneous rate $u$ into the output link. Hereafter, we simply refer to the instantaneous rate as *rate*, unless otherwise stated. Because of the impact of the cross-traffic in the network path, in the receiver, the receiving time interval between the consequent probe packets may be changed. By measuring the time interval between the received probe packets, the receiver is able to obtain the probe rate, $r$. For cases where $u \leq C - y$ no congestion is experienced, thus $r = u$. However, in cases where $u > C - y$, network is in an overload status. Therefore [8],

$$\frac{u}{r} = \max\left(1, \frac{u+y}{C}\right) = \begin{cases} 1 & u \leq C - y \\ \frac{1}{C}u + \frac{y}{C} & u > C - y \end{cases}. \quad (3)$$

The inter packet strain parameter, $\varepsilon$, is then defined as [13]

$$\varepsilon = \frac{u}{r} - 1. \quad (4)$$

Therefore, for $u > C - y$

$$\varepsilon = \frac{1}{C}u + \frac{y-C}{C}. \quad (5)$$

By setting

$$\alpha = \frac{1}{C}; \quad \beta = \frac{y-C}{C}, \quad (6)$$

the inter packet strain parameter in (4) is obtained as,

$$\varepsilon = \alpha u + \beta. \quad (7)$$

Based on the AB definition, the value of $u$ at which $\frac{u}{r}$ starts to deviate from unity is an estimate of AB. This definition of AB can be interpreted as follows: if sending rate is smaller than the AB, the packets do not cause congestion on the network path and the transmitting and receiving rates are equal. Otherwise, the packets are backlogged, thus the congestion is experienced according to (7).

Note that the fluid flow model is an asymptotic model for an actual packet transmission scenario; therefore, (7) expresses the asymptotic relation between the transmission probe rate, $u$ and the inter-packet strain, $\varepsilon$ [15].

Following the same argument for several concatenated links, it was shown in [8] that the AB of the entire path from the transmitter to the receiver can be estimated by investigating $\frac{u}{r}$ and determining the point at which $\frac{u}{r}$ starts deviating from unity. The corresponding obtained value of $u$ at this point is considered as an estimate of the path AB.

## III. AB ESTIMATION USING MULTI-RATE PROBING

By utilizing the sequence of single rate for sampling the network path, in some cases ($u \leq C - y$) we exert the load of probe packet to the path without obtaining any information about AB [13]. In this situation, another probing sequence should be sent; therefore the system time and bandwidth resource have been consumed. Since, if the probing rate can be varied in each sequence, we have further chance to obtain a sample of AB using each sequence. Although, variation of probing rate in each sequence increases the variance of probing sequences strain and degrades the statistical precision. In the consequent section we explain that how to handle this problem by efficient and simple method.

### A. Probing Procedure

In this paper, we propose a new method which employs multi-rate probing sequence as it is shown in Fig. 1. Employing multi-rate probe packet transmission enables us to probe the network path over several rates in each probing sequence.

Let $M$ be the total number of the packets in each probing sequence, and $S$ be the packet size (in bytes). The probing sequence consists of $P$ portions, each containing $(M-1)/P$ probe packets. In each portion the inter-packet times are equal and considered so that the actual probe packet transmission rate is $u_p$ $(p=1,2,\ldots,P)$. We refer to each two consequent probe packets as a *probe pair*; therefore, probe pair $i$ refers to the $i^{th}$ two consequent probe packets in a sequence.

### B. Estimation by Kalman Filtering

To estimate the available bit-rate of a path, a sequence of the above mentioned probe packets is injected into the path of interest. Then, the inter-packet strain $\varepsilon_i$, will be obtained for each probe pair $i$ $(i=1,2,\ldots,M-1)$, by measuring the inter-arrival time of the consequent packets at the receiver,

$$\varepsilon_i = \frac{(g_O)_i}{(g_I)_i} - 1, \qquad (8)$$

where, $(g_I)_i$ and $(g_O)_i$ are the initial inter-packet time for probing pair $i$ at the transmitter, and the inter-arrival time between the packets of probing pair $i$ received at the receiver side, respectively.

After receiving the last packet of the probe sequence, and obtaining the strain of all probing pairs, we obtain an AB estimation using Kalman filtering. In order to use Kalman filter, similar to [13] we model the system by the following linear equation,

$$\mathbf{x}_k = f(\mathbf{x}_{k-1}) + \mathbf{w}_{k-1} \qquad (9)$$

where $\mathbf{x}_k$ is the *state vector* at the $k^{th}$ step of the estimation which is defined as,

$$\mathbf{x}_k = [\alpha_k, \beta_k]^T, \qquad (10)$$

$f(\cdot)$ is a known function of $\mathbf{x}_{k-1}$, and $\mathbf{w}_{k-1}$ is the process noise at the $k^{th}$ step. We also define the measurement equation,

$$\mathbf{z}_k = h(\mathbf{x}_k) + \mathbf{v}_k \qquad (11)$$

where $h(\cdot)$ is a known function of $\mathbf{x}_k$, and $\mathbf{v}_k$ is the measurement noise. In this equation, $\mathbf{z}_k$ is a $P\times 1$ vector of measured strains:

$$\mathbf{z}_k = [(z_1)_k\ (z_2)_k\ \ldots (z_P)_k]^T \qquad (12)$$

where $(z_p)_k$ is the average of the measured strains of probe packets in portion $p$, at $k^{th}$ step of estimation.

Assuming an unchanged network setting in the observation interval $f(\cdot)$ can be defined as,

$$f(\mathbf{x}_{k+1}) = \mathbf{A}\mathbf{x}_k, \qquad (13)$$

where, transition matrix $\mathbf{A}$ is equal to the unit matrix $\mathbf{I}$. In addition, we consider $h(\cdot)$ as

$$h(\mathbf{x}_k) = \mathbf{H}_k \mathbf{x}_k. \qquad (14)$$

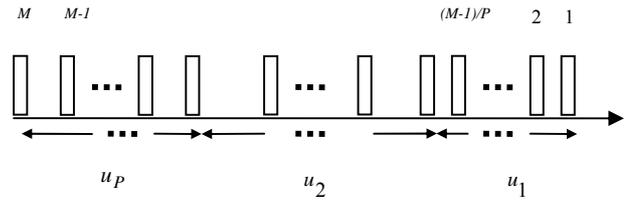

Fig. 1. MR-BART probing sequence.

In this equation, $\mathbf{H}_k$ is

$$\mathbf{H}_k = \begin{bmatrix} (u_1)_k & 1 \\ (u_2)_k & 1 \\ \vdots & \vdots \\ (u_P)_k & 1 \end{bmatrix}, \qquad (15)$$

where $(u_p)_k$ is the rate of probe packets in portion $p$, at the $k^{th}$ step of estimation.

Considering the above definitions and assumptions, we get

$$\mathbf{x}_k = \mathbf{x}_{k-1} + \mathbf{w}_{k-1} \qquad (16)$$
$$\mathbf{z}_k = \mathbf{H}_k \mathbf{x}_k + \mathbf{v}_k \quad . \qquad (17)$$

In our model we assume that

$$\mathbf{v}_k \sim \mathcal{N}(0,\mathbf{R})\ ;\ \mathbf{w}_k \sim \mathcal{N}(0,\mathbf{Q})$$

and $\mathcal{N}(0,\boldsymbol{\theta})$ denotes a zero mean Gaussian random variable with covariance matrix $\boldsymbol{\theta}$, $\mathbf{Q}$ and $\mathbf{R}$ are the covariance matrixes of the process and measurement noise, respectively. As mentioned in the above section, varying the probing rate in each probing sequence results in increasing the variance of the strain of the sequence. For handling this problem we compute the strain of each portion separately, and define the covariance matrix of probing sequence strain, $\mathbf{R}$ ($P\times P$ matrix), as

$$\mathbf{R} = \begin{bmatrix} (R_1)_k & 0 & \ldots & 0 \\ 0 & (R_2)_k & & 0 \\ \vdots & & \ddots & \vdots \\ 0 & 0 & \ldots & (R_P)_k \end{bmatrix}, \qquad (18)$$

where $(R_p)_k$ is the variance of the measured strain of probe packets in portion $p$, at the $k^{th}$ step of estimation.

Note that, in practice, the cross-traffic is not fluid flow with constant rate because of its bursty nature. Deviation from fluid flow model due to the burstiness of the cross-traffic is taken care of by considering the noise process $\mathbf{w}_k$ in (16).

To apply Kalman filtering method in (16) we need to find an appropriate value for $\mathbf{Q}$ which is a $2\times 2$ symmetric matrix. This matrix which describes the intrinsic fluctuations in the system and is related to the cross-traffic temporal fluctuation. For $\alpha$ and $\beta$ which are independent random variables, matrix $\mathbf{Q}$ can take the following simple form:

$$\mathbf{Q} = \begin{bmatrix} \lambda & 0 \\ 0 & \lambda \end{bmatrix}, \qquad (19)$$

where $\lambda$ is an adjustable parameter which is related to the cross-traffic statistics.

In our proposed approach, the Kalman filter takes $\mathbf{H}_k$, $\mathbf{z}_k$

and $\hat{\mathbf{x}}_k$ as inputs, and then obtains a new state vector $\hat{\mathbf{x}}_{k+1}$. Using the new state vector $\hat{\mathbf{x}}_{k+1}$ an estimation of the AB is then obtained as

$$\hat{A} = -\frac{\hat{\beta}}{\hat{\alpha}}. \tag{20}$$

## IV. PERFORMANCE ANALYSIS

### A. AB Estimation Error

In this section, we derive an analytical model for estimation error based on the system parameters $M$ and $P$. Here, we consider the mean square error (MSE) of our estimation

$$\xi = E(A - \hat{A})^2 = \lim_{N \to \infty} \frac{1}{N} \sum_{k=1}^{N} (A_k - \hat{A}_k)^2, \tag{21}$$

where $A_k$ refers to the AB of the bottleneck link obtained by the $k^{th}$ probing sequence. Here, $N$ is the number of probing sequences which are injected into the network path. Equality in the second line of (21) holds due to the *weak law of large numbers* [20].

Utilizing (1), we can write,

$$A = C - y_{\delta_T}(t), \tag{22}$$

where $C$ and $A$ are the capacity of bottleneck link, and AB of the path, respectively. In this equation, $y_{\delta_T}(t)$ is the average cross-traffic rate in the time interval $[t, t+\delta_T]$. In this time interval, $\delta_T$ is the inter-packet time between the first and the last packet of the probing sequence (which we refer to as *observation time* of probing sequence), and $t$ is the arrival time of the first packet in a probing sequence to the bottleneck link. If $\delta_p$ is considered as observation time of $p^{th}$ portion of probing sequence, then it can be defined as,

$$\delta_p = \frac{(M-1)S}{Pu_p}. \tag{23}$$

Total observation time of probing sequence with respect to the observation time of its portions is as follows,

$$\delta_T = \sum_{p=1}^{P} \delta_p. \tag{24}$$

Utilizing (22), we can write,

$$\hat{A} = \hat{C} - \hat{y}_{\delta_T}(t), \tag{25}$$

where $\hat{A}$, $\hat{C}$ and $\hat{y}_{\delta_T}(t)$ are the estimation of $A$, $C$ and $y_{\delta_T}(t)$ respectively. Using (22) and (25) in (21), we get,

$$\xi = E\left((C - \hat{C}) - (y_{\delta_T}(t) - \hat{y}_{\delta_T}(t))\right)^2. \tag{26}$$

Here, we assume that the capacity of the bottleneck link is known. Under such assumption the estimation problem is reduced to finding an interpretation of the following equation with respect to the system parameters,

$$\xi = E(y_{\delta_T}(t) - \hat{y}_{\delta_T}(t))^2. \tag{27}$$

By applying the above assumption to the measurement equation (17), we get,

$$\begin{bmatrix} (z_1)_k - \frac{1}{C}(u_1)_k \\ (z_2)_k - \frac{1}{C}(u_2)_k \\ \vdots \\ (z_P)_k - \frac{1}{C}(u_P)_k \end{bmatrix} = \begin{bmatrix} 1 \\ 1 \\ \vdots \\ 1 \end{bmatrix} \beta_k + \begin{bmatrix} (v_1)_k \\ (v_2)_k \\ \vdots \\ (v_P)_k \end{bmatrix}, \tag{28}$$

or equivalently,

$$\widetilde{\mathbf{z}}_k = \mathbf{h}_k \widetilde{x}_k + \mathbf{v}_k. \tag{29}$$

In the above equation, $p^{th}$ element of $\widetilde{\mathbf{z}}_k$ ($P \times 1$ vector) is as follows,

$$(\widetilde{z}_p)_k = (z_p)_k - \frac{1}{C}(u_p)_k, \tag{30}$$

and $\mathbf{h}_k$ ($P \times 1$ vector) is,

$$\mathbf{h}_k = \begin{bmatrix} 1 \\ 1 \\ \vdots \\ 1 \end{bmatrix}. \tag{31}$$

By considering the bottleneck link capacity as a known parameter, the state vector of our problem is reduced to a scalar as, $\widetilde{x}_k = \beta_k$. Therefore, (16) changes as follows,

$$\widetilde{x}_k = \widetilde{x}_{k-1} + w_{k-1}, \tag{32}$$

where $w_{k-1}$ is a Gaussian variable with zero mean and variance $\lambda$.

In our model, the noise covariance is a diagonal matrix, and thus that the components of $\mathbf{v}$ are uncorrelated. For such condition, it is advantageous to consider the components of $\mathbf{z}$ as independent scalar measurements, rather than as a vector measurement in Kalman filter [16].

Based on [16], if we consider the components of $\mathbf{z}$ as independent scalar measurements, then the filter implementation requires $P$ iterations. The updating process can be implemented iteratively using the rows of $\mathbf{H}$ as the measurement matrices (with row dimension equal to 1) and the diagonal elements of $\mathbf{R}$ as the corresponding (scalar) measurement noise covariance as the following equations,

$$\mathbf{k}_k^{[p]} = \frac{1}{(\mathbf{h}_p)_k \mathbf{\psi}_k^{[p-1]}(\mathbf{h}_p)_k^T + (R_p)_k} \mathbf{\psi}_k^{[p-1]}(\mathbf{h}_p)_k^T \tag{33}$$

$$\mathbf{\psi}_k^{[p]} = \mathbf{\psi}_k^{[p-1]} - \mathbf{k}_k^{[p]}(\mathbf{h}_p)_k \mathbf{\psi}_k^{[p-1]} \tag{34}$$

$$\hat{\mathbf{x}}_k^{[p]} = \hat{\mathbf{x}}_k^{[p-1]} + \mathbf{k}_k^{[p]}\left[\mathbf{z}_k^{[p]} - (\mathbf{h}_p)_k \hat{\mathbf{x}}_k^{[p-1]}\right] \tag{35}$$

for $p = 1, 2, \ldots, P$, using the initial values $\mathbf{\psi}_k^{[0]}$ and $\hat{\mathbf{x}}_k^{[0]}$. Where $\mathbf{\psi}_k^{[P]}$ is state estimate error covariance matrix, $\hat{\mathbf{x}}_k^{[P]}$ is the estimated state vector for $p^{th}$ portion of $k^{th}$ probing sequence, and $\mathbf{k}$ is the *Kalman gain*. The intermediate variable $(R_p)_k$ is the $p^{th}$ diagonal element of the $P \times P$ diagonal matrix $\mathbf{R}_k$ and $(\mathbf{h}_p)_k$ is the $p^{th}$ row of the $P \times 2$ matrix $\mathbf{H}_k$. The final value $\hat{\mathbf{x}}_k^{[p]}$ i.e., $\hat{\mathbf{x}}_k^{[P]}$ is an estimate of the state vector obtained from sending one probing sequence consists of $P$ portions.

Combining new definition of the parameters of KF equations with equations (33-35) result in,

$$k_k^{[p]} = \frac{1}{\psi_k^{[p-1]} + (R_p)_k} \psi_k^{[p-1]}, \quad (36)$$

$$\psi_k^{[p]} = (1 - k_k^{[p]})\psi_k^{[p-1]}, \quad (37)$$

$$\hat{\tilde{x}}_k^{[p]} = \hat{\tilde{x}}_k^{[p-1]} + k_k^{[p]}[\mathbf{z}_k^{[p]} - \hat{\tilde{x}}_k^{[p-1]}]. \quad (38)$$

Utilizing the KF equations yields

$$\psi_k = E(\tilde{x}_k - \hat{\tilde{x}}_k)^2, \quad (39)$$

or equivalently,

$$\psi_k = E(\beta_k - \hat{\beta}_k)^2. \quad (40)$$

Recall that,

$$\psi_k = \psi_k^{[P]}. \quad (41)$$

From (6) and (40) we have,

$$\psi_k = E\left(\frac{(y_{\delta_T}(t) - \hat{y}_{\delta_T}(t))_k}{C}\right)^2. \quad (42)$$

Combining (42) and (21) we get,

$$\xi = \lim_{N \to \infty} \frac{C^2}{N} \sum_{k=1}^{N} \psi_k. \quad (43)$$

If we can find the value of $(R_p)_k$ analytically, then using initial value for $\psi_k^{[0]}$ the relationship between $\xi$ and system parameters will be obtained. From [15] we have,

$$\begin{cases} (g_O)_p = \frac{(g_I)_p y_{\delta_p}(t_p)}{C} + \frac{S}{C} & (g_I)_p \leq \frac{S}{C} \\ \frac{(g_I)_p y_{\delta_p}(t_p)}{C} + \frac{S}{C} \leq (g_O)_p \leq \frac{(g_I)_p y_{\delta_p}(t_p)}{C} + (g_I)_p & (g_I)_p > \frac{S}{C} \end{cases} \quad (44)$$

$$\frac{D_{\delta_p}(t_p)}{(M-1)/P} + (g_I)_p \leq (g_O)_p \leq \frac{D_{\delta_p}(t_p)}{(M-1)/P} + (g_I)_p + \frac{S}{C}, \quad (45)$$

where $(g_I)_p$ and $(g_O)_p$ are the average inter-packet time of $p^{th}$ portion at the transmitter, and the receiver, respectively. For defining the $D_{\delta_p}(t_p)$ in (45), we first define the *hop-workload* process $W(t)$, as sum of service time of all packets in the queue and the remaining service time of the packet in service [15]. By this definition, $D_{\delta_p}(t_p)$ is defined as the difference between hop-workload at time $t_p$ and $(t_p + \delta_p)$, i.e.,

$$D_{\delta_p}(t_p) = W(t_p + \delta_p) - W(t_p) \quad (46)$$

Rearranging the (44) and (45), we get the strain of $p^{th}$ portion as,

$$\begin{cases} \frac{(g_O)_p}{(g_I)_p} - 1 = \frac{y_{\delta_p}(t_p)}{C} + \frac{S}{(g_I)_p C} - 1 & (g_I)_p \leq \frac{S}{C} \\ \frac{y_{\delta_p}(t_p)}{C} + \frac{S}{(g_I)_p C} - 1 \leq \frac{(g_O)_p}{(g_I)_p} - 1 \leq \frac{y_{\delta_p}(t_p)}{C} & (g_I)_p > \frac{S}{C} \end{cases}, \quad (47)$$

$$\frac{D_{\delta_p}(t_p)}{((M-1)/P)(g_I)_p} \leq \frac{(g_O)_p}{(g_I)_p} - 1 \leq \frac{D_{\delta_p}(t_p)}{((M-1)/P)(g_I)_p} + \frac{S}{(g_I)_p C}. \quad (48)$$

In order to find the bounds of strain, we analyze the behavior of $D_{\delta_p}(t_p)$. We can write $W(t_p + \delta_p)$ as follows,

$$W(t_p + \delta_p) = W(t_p) + \frac{b(t_p + \delta_p) - b(t_p)}{C} - \delta_p + I_{\delta_p}(t_p), \quad (49)$$

where $b(t)$ is the total volume of cross-traffic (measured in bits) receive by the bottleneck link up to time instant $t$. Using this definition we can write the $y_{\delta_p}(t_p)$ as,

$$y_{\delta_p}(t_p) = \frac{b(t_p + \delta_p) - b(t_p)}{\delta_p}. \quad (50)$$

In addition, $I_{\delta_p}(t_p)$ is the total amount of idle time of the link in the time interval $[t_p, t_p + \delta_p]$ [15]. If the link is busy transmitting the cross-traffic packets in the all time of this time interval then $I_{\delta_p}(t_p) = 0$. If in this time interval link is idle $I_{\delta_p}(t_p) = \delta_p$, thus,

$$0 \leq I_{\delta_p}(t_p) \leq \delta_p. \quad (51)$$

Combining (46), (49) and (50) we get,

$$D_{\delta_p}(t_p) = \frac{\delta_p}{C} y_{\delta_p}(t_p) - \delta_p + I_{\delta_p}(t_p). \quad (52)$$

Applying the bounds of $I_{\delta_p}(t_p)$ on (52) results in,

$$\frac{\delta_p}{C} y_{\delta_p}(t_p) - \delta_p \leq D_{\delta_p}(t_p) \leq \frac{\delta_p}{C} y_{\delta_p}(t_p). \quad (53)$$

Combining (48) and (53), we get,

$$\frac{y_{\delta_p}(t_p)}{C} - 1 \leq \frac{(g_O)_p}{(g_I)_p} - 1 \leq \frac{y_{\delta_p}(t_p)}{C} + \frac{S}{(g_I)_p C}. \quad (54)$$

Collecting (47) and (54) leads to,

$$\begin{cases} \frac{(g_O)_p}{(g_I)_p} - 1 = \frac{y_{\delta_p}(t_p)}{C} + \frac{S}{(g_I)_p C} - 1 & (g_I)_p \leq \frac{S}{C} \\ \frac{y_{\delta_p}(t_p)}{C} - 1 \leq \frac{(g_O)_p}{(g_I)_p} - 1 \leq \frac{y_{\delta_p}(t_p)}{C} + \frac{S}{(g_I)_p C} & (g_I)_p > \frac{S}{C} \end{cases}. \quad (55)$$

In each portion of probing sequence, $C$, $S$ and $(g_I)_p$ are constant; therefore, the variance of the above equation is

$$\text{Var}\left(\frac{(g_O)_p}{(g_I)_p} - 1\right) = \frac{1}{C^2} \text{Var}(y_{\delta_p}(t_p)), \quad (56)$$

or equivalently,

$$R_p = \frac{1}{C^2} \text{Var}(y_{\delta_p}(t_p)). \quad (57)$$

where $R_p$ is the variance of the measured strain of probe packets in portion $p$.

Recall that in our model the cross-traffic is self similar which generates from fBm distribution, so we can write [19],

$$b(t) = \mu t + \sigma \omega(t), \quad (58)$$

where $\mu$ is the average rate, $\sigma$ is a controlling factor of fluctuation, and $\omega(t)$ is the fractional Brownain motion (fBm) process describing the cross-traffic. fBm process $\omega(t)$ is a Gaussian process with zero mean, which is stationary increment and its covariance function is [18],

$$E(\omega(t+\tau)\omega(t)) = \frac{1}{2}\left(|t+\tau|^{2H} + |t|^{2H} - |\tau|^{2H}\right). \quad (59)$$

where $H$ denotes the *self-similarity index*.

From (50) and (59) we get,

$$\text{Var}(y_{\delta_p}(t_p)) = \sigma^2 \, \delta_p^{2H-2} . \quad (60)$$

By replacing the above equation in to the (57) we have,

$$R_p = \frac{1}{C^2}\left(\sigma^2 \, \delta_p^{2H-2}\right). \quad (61)$$

Utilizing (61), (36), (37) and (43), we will have,

$$\xi = \lim_{N \to \infty} \frac{C^2}{N} \sum_{k=1}^{N} \psi_k^{[P-1]} - \frac{\left(\psi_k^{[P-1]}\right)^2}{\psi_k^{[P-1]} + \frac{1}{C^2}\left(\sigma^2 \delta_P^{2H-2}\right)}. \quad (62)$$

Therefore, we obtained $\xi$ with respect to the probing sequence parameters and features of cross-traffic.

In order to analyze the behavior of *MSE* with respect to the probing sequence parameter, we plot the *MSE* versus $P$ and $M$ in Figs. 2, and 3, respectively. In addition, in these figures a comparison between simulation and analytical results is made simultaneously. The simulated network has one bottleneck link as shown in Fig. 4. The bottleneck of the network path is the link between the two considered routers. In our model, the bottleneck link has the minimum capacity along the network path of interest and its capacity is $C = 1$ (bit/s). For the cross-traffic, we consider a fractional Brownian motion (fBm) model with $H = 0.7$ [17]. In addition, the number of probing sequences which are injected into the network path considered as $N = 1000$.

Fig. 2 shows the impact of the increasing $P$ on the accuracy of estimation. As it is seen in this figure, increasing the $P$ result in decreasing *MSE* both in the simulation and theoretical results. By increasing $P$ the dimension of the measurement vector of Kalman filter will increase and therefore we can obtain a more accurate estimation. Although, the slop of the decreasing the *MSE* in theoretical results is larger than that of simulation. The reason of this phenomenon will be discussed later.

Fig. 3 shows the impact of the increasing $M$ on the behavior of *MSE*. For all $P$, *MSE* decreases as the number of probe packets increase. This is because of the increasing the observation interval (due to the increasing the $M$) and therefore obtaining an appropriate estimate for the average cross-traffic arrival rate.

The experimental results we obtained agree with our analytical findings. Although, the slop of decreasing the *MSE* of analytical results are restively larger than that of simulation. The interpretation of this phenomenon is as follows. As mentioned in section II, the fluid flow model is an asymptotic model for an actual packet transmission scenario; therefore (7) expresses the asymptotic relation between the transmission probe rate, $u$ and the inter-packet strain, $\varepsilon$. In the actual and therefore in the simulated network, we have bursty arrivals of discrete cross-traffic packets.

Therefore, the relationship between $\varepsilon$ and $u$ deviate from (7). The amount of this deviation is influenced by the probing sequence parameters [15]. As the $P$ increases, the variation of probe packet rate in each sequence will increase (increasing the burstiness). On the other hand since the deviation of fluid flow model is originated from the burstiness of the traffic, therefore the deviation from fluid flow model will increase. This effect decreases the accuracy of the network modeling. However, it should be noted that, by increasing the $P$ accuracy of the KF estimation will increase. Based on our observation, increasing $P$, at first increases the accuracy of estimation (by increasing the accuracy of KF) but farther increasing the $P$ cause to degrading the accuracy of modeling and influence the accuracy of estimation. Since in our theoretical analysis deviation from fluid flow model due to the burstiness of the cross-traffic is only taken care of by considering the process noise, therefore the effect of increasing the KF accuracy by increasing $P$ is more obvious than degrading the deviation of modeling. Base on this reason, by increasing $P$ the *MSE* of the analytical curves will decrease faster than curves of simulation.

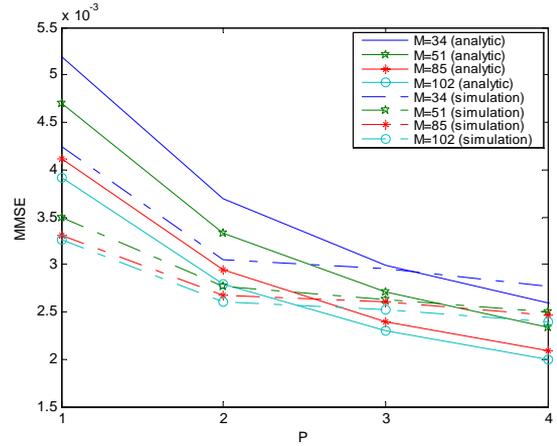

Fig. 2. Impact of varying $P$ on the accuracy of MR-BART for fixed values of $M$ ($C = 1$ bit/s, $H = 0.7$).

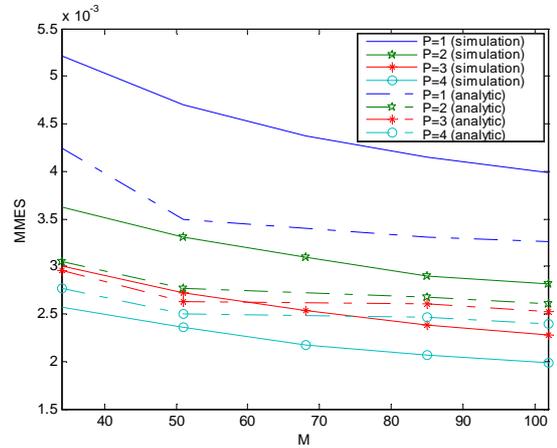

Fig. 3. Impact of varying $M$ on the accuracy of MR-BART for fixed values of $P$ ($C = 1$ bit/s, $H = 0.7$).

### B. Impact of P on the Complexity of Computation

As mentioned above, in our model the noise covariance is a diagonal matrix, and it results that the components of $\mathbf{v}$ are uncorrelated. For such condition, we consider the components of **z** as independent scalar measurements.

According to [16], the number of operations which is required for processing **z** (a vector by dimension $P \times 1$) as $P$ successive scalar measurements is significantly less than that of the corresponding number of operations for vector measurement processing. It is shown that the complexity of computations for the vector implementation of Kalman filter equations grows as $P^3$ [16], whereas that of the scalar implementation these equations grows only as $P$. Furthermore, if we consider the components of **z** as independent scalar measurements, we can avoid matrix inversion in the implementation of the Kalman filter equations and improve the robustness of the computation against error [16]. Therefore, by processing vector **z** as $P$ successive scalar measurements we can reduce the computation time comparing to processing based on **z** as a vector and improves numerical accuracy.

By utilizing above implementation, we can reduce the complexity of computation, so that, in our method the complexity of computation increase as $P$ (instead of $P^3$) rather than BART.

## V. SIMULATION RESULTS

We consider a general model for a network path with one bottleneck link which is shown in Fig. 4. The bottleneck of the network path is the link between the two considered routers. In our model, the bottleneck link has the minimum capacity along the network path of interest. A similar model is also considered in [13] for a network path with one bottleneck link.

For the cross-traffic, in this paper, we consider self similar traffic which is explored from a fractional Brownian motion (fBm) model with $H = 0.7$ [17]. This model is shown to be a good fit to the aggregated traffic in a data network with bursty traffic sources [18].

All links in the simulated network have a nominal capacity of 100 (Mbits/s), except the bottleneck link between the two routers, which has the capacity of 10 (Mbits/s). The cross-traffic which is used in our simulations is the output of a fractional Brownian motion model [17].

MR-BART was configured to produce an estimate every second, i.e. the inter-departure time between two consecutive probing sequences is one second. In our simulations, we set $N = 1000$ and $MSE$ is normalized to the square of bottleneck link capacity [13]. Furthermore, in our simulations we find that the accuracy of estimation will not change significantly for $S > 1500$ (bytes), hence we set $S = 1500$ (bytes) in all experiments unless otherwise stated.

Fig. 5 shows the AB estimation utilizing the MR-BART method. In this figure, in order to estimating AB, we use probing sequence with length $M = 34$ and $P = 2$. As it is seen in this figure, we obtain reasonable level of accuracy of AB estimation by using tow portion in each probing sequence. This figure shows that our method is able to track both slow and fast variation of cross-traffic rate simultaneously.

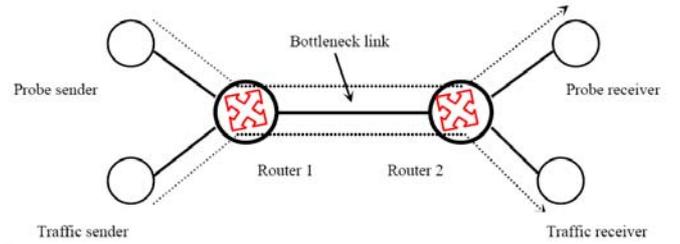

Fig. 4. A schematic view of the measurement test-bed.

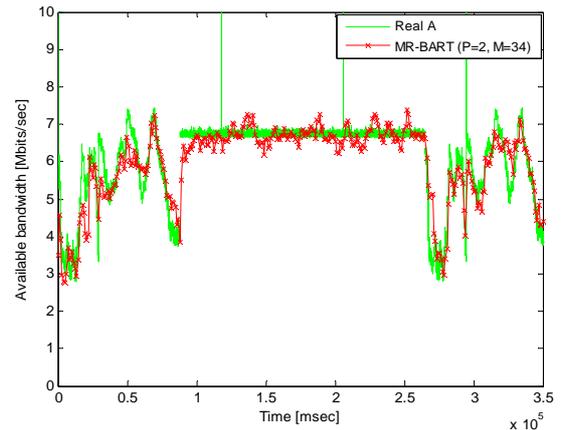

Fig. 5. MR-BART estimation of AB ($P = 2, M = 34$)

### A. Impact of M and S on the Estimation Accuracy

Fig. 6 depicts the AB estimation by using the proposed method in different values of $M$, and $S$. In this figure, we set $P = 3$. As seen in this figure, increasing $M$ and $S$ leads to a better estimation of AB. This is because of the increasing the observation interval (due to the increasing the $M$ and $S$) and therefore obtaining an appropriate estimate for the average cross-traffic arrival rate. In Table I, the $MSE$ of estimation has been reported. In our experiments we investigate the behavior of $MSE$ versus $M$ and $S$, and find that larger $M$ and $S$, result in smaller $MSE$. In addition we observe that the accuracy of estimation for $S > 1500$ bytes and $M > 34$ will not change significantly. Therefore, hereafter we use $S = 1500$ bytes in our simulations.

### B. Impact of M and P on the Estimation Accuracy

Here we study the impact of parameters $P$ and $M$, on the accuracy of the AB estimation. As it is shown in Fig. 7, the simulation result confirms that, increasing $P$ in small values of $M$ leads to decreasing the accuracy of estimation. But if we choose $M$ large enough, then selecting the large value of $P$, results in accurate estimation of AB. From Fig. 7 we see that for $M = 17$, $P = 3$ shows better performance rather than $P = 4$. But if we select $M$ large enough ($M = 34$), then $P = 4$ results in better estimation of AB.

As mentioned above, if we choose $M$ large enough then increasing $P$ results in more accurate estimation of AB. However, our simulation results show that, for $P \geq 5$, error of estimation will not change significantly by increasing $P$. The results of simulation have been reported in Table II.

### C. Modeling of Error Based on the System Parameters

In this subsection, we derive an empirical model for estimation error based on the system parameters $M$ and $P$. We consider a general model for a network path with one bottleneck link which is shown in Fig. 4. The bottleneck of the network path is the link between the two considered routers. In our model, the bottleneck link has the minimum capacity along the network path of interest.

In the above scenario, we obtain $\xi$ from (21) using $16 \leq M \leq 100$ probe packets, and $1 \leq P \leq 5$ portions. Based on our observations from simulation results (see Figs. 8, 9, 10 and 11), we obtain the following expression by curve fitting on the simulation results of $\xi$ based on $M$ and $P$ as follows:

$$\xi = a \frac{e^{1.1P}}{M^b (P^2 + P)}, \quad (63)$$

where $a$ and $b$ are two positive parameters. The values of $a$ and $b$ depend on $P$ and the bottleneck capacity, as shown in Table III. In Table III, we report suitable values of $a$ and $b$ for $P = 1, 2, \ldots, 5$, and bottleneck capacity $10 \leq C \leq 70$ (Mbits/s). We show later that for values of $P$ greater than 5, we do not achieve significant improvement on the estimation performance.

Here after we analyze the impact of parameters $M$ and $P$ on the accuracy of the proposed AB estimation method. Based on (63), $\xi$ is a function of two parameters, $M$ and $P$. The partial derivative of $\xi(P,M)$ with respect to $M$ is,

$$\frac{\partial \xi}{\partial M} = -ab \frac{e^{1.1P} M^{b+1}}{(P^2 + P)}. \quad (64)$$

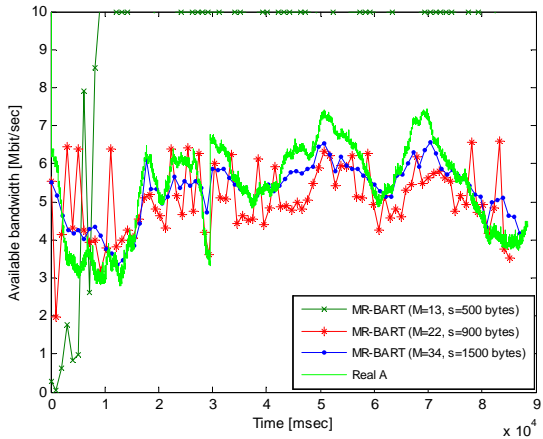

Fig. 6. Impact of varying $M$ and $S$ on accuracy of MR-BART estimate ($P = 3$).

TABLE I
IMPACT OF VARIATION $M$ AND $S$ ($P=3$) ON ACCURACY OF MR-BART ESTIMATION.

|  | $S$ = 500 bytes $M$ =13 | $S$ = 900 bytes $M$ =22 | $S$ = 1500 bytes $M$ =34 |
|---|---|---|---|
| MSE | 0.190 | 0.013 | 0.009 |

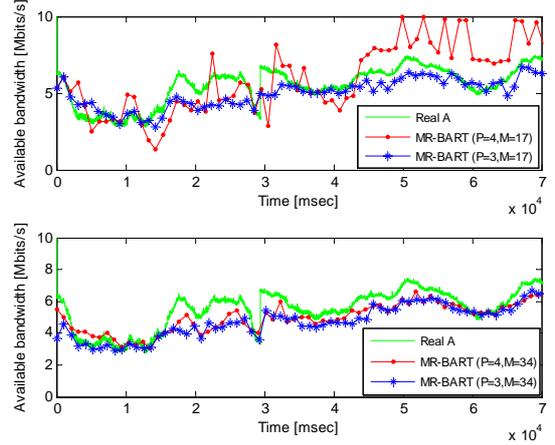

Fig. 7. Impact of varying $P$ and $M$ on accuracy of MR-BART estimate ($S = 1500$ bytes).

TABLE II
MSE OF MR-BART FOR $P \geq 5$ ($S = 1500$ bytes).

| P \ M | 5 | 6 | 7 | 8 |
|---|---|---|---|---|
| 34 | 0.009 | 0.009 | 0.010 | 0.011 |
| 45 | 0.008 | 0.009 | 0.009 | 0.010 |
| 100 | 0.006 | 0.006 | 0.006 | 0.005 |

TABLE III
VALUES OF $a$ AND $b$ IN DIFFERENT $P$ AND $C$

| P \ C (Mbits/s) | 1 | 2 | 3 | 4 | 5 |
|---|---|---|---|---|---|
| 10 | $a$=0.04 $b$=0.04 | $a$=0.06 $b$=0.33 | $a$=0.01 $b$=0.33 | $a$=0.26 $b$=1.26 | $a$=0.15 $b$=1.26 |
| 30 | $a$=0.07 $b$=0.21 | $a$=0.10 $b$=0.16 | $a$=0.02 $b$=0.25 | $a$=0.08 $b$=0.63 | $a$=0.05 $b$=0.63 |
| 50 | $a$=0.10 $b$=0.08 | $a$=0.16 $b$=0.33 | $a$=0.02 $b$=0.51 | $a$=0.41 $b$=0.94 | $a$=0.41 $b$=0.94 |
| 70 | $a$=0.32 $b$=0.45 | $a$=0.29 $b$=0.53 | $a$=0.27 $b$=0.73 | $a$=0.44 $b$=1 | $a$=0.36 $b$=1.14 |

Equation (64) denotes that $\xi$ is a descending function of $M$. We also note that the minimum of $\xi$ for a constant $M$, depends on not only $P$ but also $a$ and $b$. Therefore, to obtain the minimum $\xi$ for a given $M$, we can find the suitable values of $a$ and $b$ for $P = 1, 2, \ldots, 5$, from Table III, and then utilizing (63) to find the value of $P$ for which the minimum of $\xi$ occurs.

If we rearrange the (63), then $M$ is

$$M = \sqrt[b]{\frac{a e^{1.1P}}{\xi (P^2 + P)}}, \quad (65)$$

From (65), we can obtain the appropriate value of $M$ for a given $\xi$ in a bottleneck capacity of interest, through the following steps

**1st step.** Considering a value for $P$, by regarding the constrain of computational complexity.

**2$^{nd}$ step.** Finding suitable values of $a$ and $b$ for selected $P$, using Table III.

**3$^{rd}$ step.** Computing $M$ from (65), which results in to given $\xi$.

In order to gain more insight on the behavior of $\xi$ based on $M$ and $P$, we plot $\xi(M)$ for the fixed values of $P$ in Figs. 8 and 9, and $\xi(P)$ for the fixed values of $M$ in Figs. 10 and 11 utilizing the (63) and simulation in the above scenario. In Figs. 8 and 10 we consider bottleneck link capacity $C = 10\,(\text{Mbits/s})$ and in Figs. 9 and 11, $C = 70\,(\text{Mbits/s})$.

As it is seen in Figs. 8 and 9, by increasing $M$ for a constant $P$, the estimation error $\xi$ is decreased. Comparing to the BART (where $P = 1$), in our proposed method the input vector of Kalman filter has a larger dimension ($P > 1$). Therefore, we expect higher estimation accuracy than that of obtained by the conventional BART method, which is confirmed by the simulation results in Figs. 8 and 9.

In Figs. 10 and 11, the behavior of $\xi$ for different values of $M$ has been plotted. As it is seen, $\xi$ for a constant $M$ has a minimum versus $P$. That is, if $M$ remains unchanged and $P$ increases, then $\xi$ decreases to its minimum value but thereafter, with increasing the $P$, $\xi$ will increase. It should be noted that, this problem is more obvious for small values of $M$. If we choose a large enough $M$ (i.e., $M \geq 34$), then the increase in $\xi$ beyond its minimum point can be ignored. The interpretation of this phenomenon is as the following. As mentioned in previous section, by increasing $P$ the dimension of the measurement vector of Kalman filter is increased therefore, we can obtain a more accurate estimation. But, if the increase in the estimation accuracy is obtained for a constant $M$, then the observation time interval for each portion of probe packets is decreased. Thus, we are not able to obtain an appropriate estimate of the average cross-traffic arrival rate in each observation time interval [15]. Consequently, the accuracy of estimation may be degraded.

Simulation results confirm that for small number of probe packets ($M \leq 45$) the minimum of $\xi$ occurs in $P = 3$. However, for $M > 45$, the minimum value of $\xi$ occurs at $P = 4$ or $P = 5$. Note that, the error estimation at $P = 4$ or $P = 5$ is marginally smaller than that of achieved for $P = 3$. Therefore, in practice we have a tradeoff between the accuracy of estimation, the cost of overload of probe packets and computational complexity due to increasing $P$.

### D. Comparison of MR-BART and BART in the Test-bed Environment

Here, we compare the performance of our method with BART method in different situations. In Fig. 12 we show the impact of increasing $M$ on the accuracy of estimation. As it is seen, the accuracy of estimation for $M > 34$ will not change significantly. But, since in the BART method $M = 17$ has been selected as the number of packet in which the accuracy of estimation is reasonable, therefore we also select this number for comparing our method with BART method.

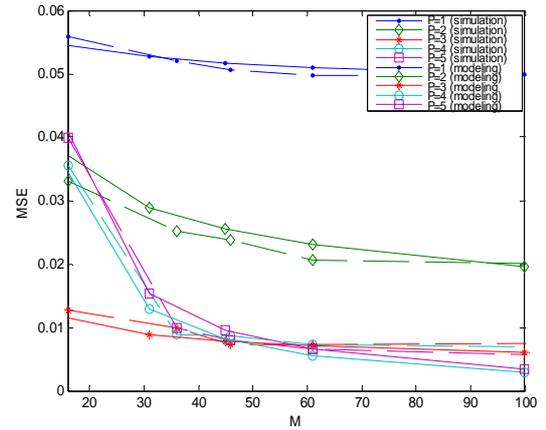

Fig. 8. Impact of varying $M$ on the accuracy of MR-BART for fixed values of $P$ ($C = 10\,\text{Mbits/s}$).

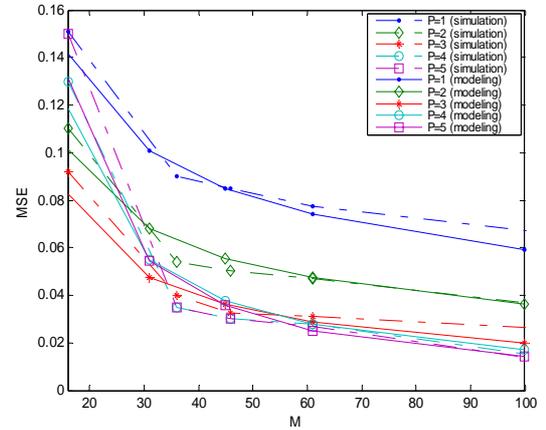

Fig. 9. Impact of varying $M$ on the accuracy of MR-BART for fixed values of $P$ ($C = 70\,\text{Mbits/s}$).

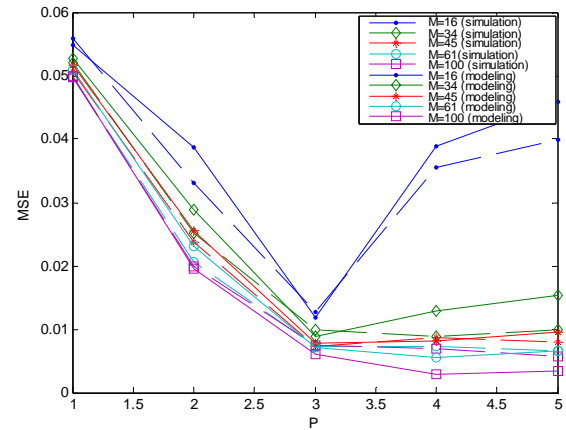

Fig. 10. Impact of varying $P$ on the accuracy of MR-BART for fixed values of $M$ ($C = 10\,\text{Mbits/s}$).

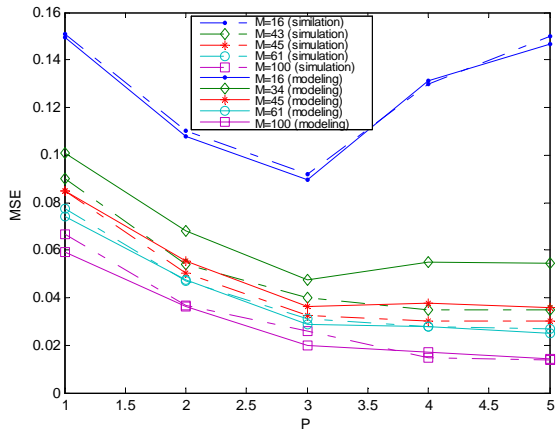

Fig. 11. Impact of varying $P$ on the accuracy of MR-BART for fixed values of $M$ ($C = 70$ Mbits/s).

Figs. 13 and 14 depict the AB versus time, and compare the behavior of MR-BART ($P = 2$ and $P = 3$ respectively) and BART when synthetic cross-traffic has been used in the test-bed. From this figure, it is quite clear that MR-BART, estimates the available bandwidth more accurately than BART. In addition, these figures show that our proposed method is able to track slow and fast variation of cross-traffic rate simultaneously. Whereas in BART method, the parameter **Q** (covariance matrixes of the noise process) should be tune for each type of fluctuation in cross-traffic.

In utilizing Kalman filter as an estimator, we need to consider an initial value for state vector. In our estimation problem, we need to initialize available bandwidth i.e., $(\hat{A})_0$ as an initial estimate of initial real value i.e. $(A)_0$. Therefore in our simulations, we select different values for $(\hat{A})_0$ and analyze the behavior of $\xi$ for inappropriate initial value of $(\hat{A})_0$. In Table IV we report the results of these experiments and compare the robustness of MR-BART and BART against the deflection from the real values of initial available bandwidth.

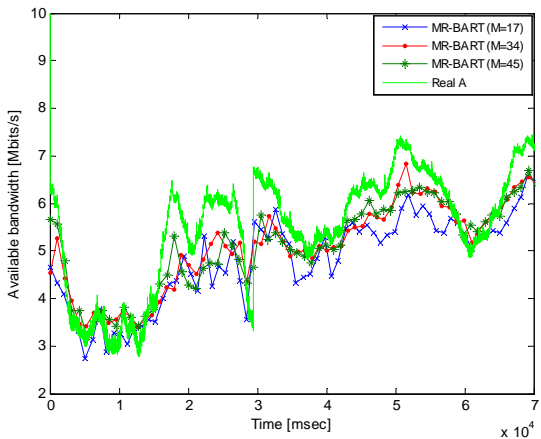

Fig. 12. Impact of variation of $M$ on accuracy of MR-BART estimate ($P = 3$, $S = 1500$ bytes).

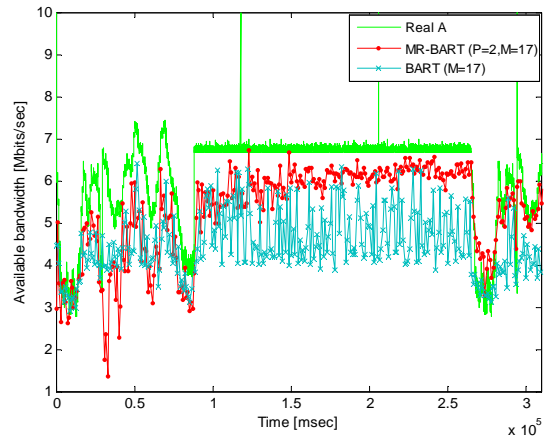

Fig. 13. Comparison of MR-BART and BART ($P = 2, M = 17, S = 1500$ bytes).

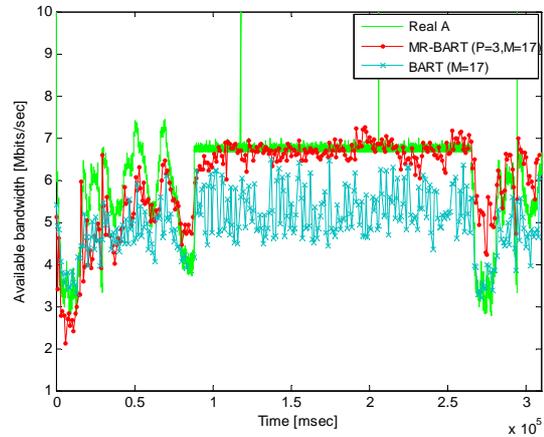

Fig. 14. Comparison of MR-BART and BART ($P = 3, M = 17, S = 1500$ bytes).

TABLE IV
$MSE$ OF MR-BART ($P = 2$) AND BART ESTIMATION WHEN USING THREE ARBITRARY INITIAL STATES FOR KALMAN FILTER
$S = 1500$ bytes, $M = 17, (A)_0 = 6.2$ Mbits/sec.

| $(\hat{A})_0$ Mbits/sec | **MR-BART** | **BART** |
|---|---|---|
| 2.5 | 0.033 | 0.057 |
| 8.5 | 0.008 | 0.022 |
| 5 | 0.006 | 0.009 |

From this Table, we see that each $(\hat{A})_0$ which is near to the $(A)_0$ leads to the smaller $MSE$ than that of is further. In addition, it can be concluded that MR-BART shows better performance in case of inappropriate initial state of Kalman filter compared to BART.

## VI. CONCLUSION

In this paper we proposed a new method called MR-BART. This method is an efficient method for real-time estimation of the available bit-rate in a network path with concurrent cross-traffic using Kalman filtering. In the proposed method a probing sequence consists of multi-rate probe packets is utilized. Indeed, in this method by using the new parameter i.e. $P$, we have more degree of freedom to design an estimator of Available Bandwidth. The proposed method is highly accurate and converges quickly in compared with conventional BART method. In addition, this method is robust against inappropriate initial value of Kalman filter. Due to the special feature of the covariance matrixes of the measurement noise, the number of computations grows only as the dimension of the measurement vector. Therefore, we can still obtain a real-time more accurate estimate of the AB than the conventional BART estimate, with marginal addition to the complexity of computation.

ACKNOWLEDGMENT

The authors would like to express their gratitude Prof. M. Nasiri-Kenari, Sharif University of Technology, Tehran, Iran, for a wealth of comments and suggestions that significantly improved both the presentation and the content of this paper.